\documentclass[10pt, aps,twocolumn,pra,superscriptaddress]{revtex4-2}

\usepackage[colorlinks=true,linkcolor=blue,citecolor=blue,urlcolor=blue]{hyperref}
\usepackage{amsmath}
\usepackage{amssymb}
\usepackage{mathtools}
\usepackage{dsfont}
\usepackage{bm}
\usepackage{cases}
\usepackage{graphicx}
\usepackage{dcolumn}
\usepackage{braket}
\usepackage{siunitx}
\usepackage{stackengine}
\usepackage[caption=false]{subfig}
\usepackage[all]{hypcap}
\usepackage[dvipsnames,table]{xcolor}
\usepackage[capitalise, nameinlink]{cleveref}
\usepackage[normalem]{ulem}
\usepackage{soul}
\crefname{appendix}{Apd.}{Apds.}
\crefname{table}{Tab.}{Tabs.}
\crefname{section}{Sec.}{Secs.}

\DeclareSIUnit\bar{bar}
\DeclareSIUnit\mbar{\milli\bar}
\newcommand{\iden}{\mathds{1}}

\newcommand{\df}{\mathrm{d}}
\newcommand{\hH}{\hat{H}}

\newcommand{\hx}{\hat{x}}

\newcommand{\hp}{\hat{p}}
\newcommand{\hO}{\hat{O}}

\newcommand{\ha}{\hat{a}}

\newcommand{\hb}{\hat{b}}

\newcommand{\hr}{\hat{\rho}}
\newcommand{\ie}{\textit{i.e.}~}
\newcommand{\eg}{\textit{e.g.}~}
\newcommand{\cf}{cf.~}

\newcommand{\calH}{\mathcal{H}}

\newcommand{\tr}[1]{\text{Tr}\left[#1\right]}
\newcommand{\panel}[1]{Panel~\textbf{#1)}}

\newcolumntype{Y}{>{\centering\arraybackslash}X}

\renewenvironment{psmallmatrix}
  {\left(\begin{smallmatrix}}
  {\end{smallmatrix}\right)}

\begin{document}

\title{Squeezing below the ground state of motion of a continuously monitored levitating nanoparticle}
\author{Q.~Wu}
\email[Corresponding author. Email: ]{qiongyuan.wu@qub.ac.uk}
\affiliation{Centre for Quantum Materials and Technologies, School of Mathematics and Physics, Queen's University Belfast, BT7 1NN, United Kingdom}
\author{D. A.~Chisholm}
\affiliation{Centre for Quantum Materials and Technologies, School of Mathematics and Physics, Queen's University Belfast, BT7 1NN, United Kingdom}
\author{R.~Muffato}
\affiliation{Department of Physics, Pontifical Catholic University of Rio de Janeiro, Brazil}
\affiliation{School of Physics and Astronomy, University of Southampton, Southampton SO17 1BJ, United Kingdom}
\author{T.~Georgescu}
\affiliation{School of Physics and Astronomy, University of Southampton, Southampton SO17 1BJ, United Kingdom}
\author{J.~Homans}
\affiliation{School of Physics and Astronomy, University of Southampton, Southampton SO17 1BJ, United Kingdom}
\author{H.~Ulbricht}
\affiliation{School of Physics and Astronomy, University of Southampton, Southampton SO17 1BJ, United Kingdom}
\author{M.~Carlesso}
\affiliation{Department of Physics, University of Trieste, Strada Costiera 11, 34151 Trieste, Italy}
\affiliation{Istituto Nazionale di Fisica Nucleare, Trieste Section, Via Valerio 2, 34127 Trieste, Italy}
\affiliation{Centre for Quantum Materials and Technologies, School of Mathematics and Physics, Queen's University Belfast, BT7 1NN, United Kingdom}
\author{M.~Paternostro}
\affiliation{Universit\`a degli Studi di Palermo, Dipartimento di Fisica e Chimica - Emilio Segr\`e, via Archirafi 36, I-90123 Palermo, Italy}
\affiliation{Centre for Quantum Materials and Technologies, School of Mathematics and Physics, Queen's University Belfast, BT7 1NN, United Kingdom}
\date{\today}

\begin{abstract}
Squeezing is a crucial resource for quantum information processing and quantum sensing. In levitated nanomechanics, squeezed states of motion can be generated via temporal control of the trapping frequency of a massive particle.
However, the amount of achievable squeezing typically suffers from detrimental environmental effects. {We propose a scheme for the generation of 
significant levels of mechanical squeezing in the motional state of a levitated nanoparticle by leveraging on the careful temporal control of the trapping potential.
We analyze the performance of such a scheme by  fully accounting for the most relevant sources of noise, including measurement backaction.} The feasibility of our proposal, which is close to experimental state-of-the-art, makes it a valuable tool for quantum state engineering.
\end{abstract}

\maketitle

Quantum sensing, which aims at achieving the efficient probing of the properties of a quantum system and through quantum resources, is a task of key relevance in applications such as thermometry~\cite{DePasquale2018}, environment characterization~\cite{benedetti_2018, bina_2018, tamascelli_2020, barr2023}, detection of gravitational waves~\cite{PhysRevLett.116.061102}, quantum illumination and quantum radars~\cite{torrome2023advances, karsa2023quantum} and being able to detect gravity-induced entanglement~\cite{bose2017spin, marletto2017gravitationally}.
To infer the information about the target system, quantum sensing uses auxiliary probing systems that can be directly controlled and measured, such that after the interaction the measurement results of the probing systems reflect the property of the target system.
By suitably engineering the initial state of the probing system, it is often possible to obtain a significant sensing advantage~\cite{Huelga1997Improvement, Ulam2001Spin, Mitchell2004Super, Leibfried2004Toward, Lee2002quantum, Genoni2011Optical, Rafal2014Using, Huang2016Usefullness}. 
Specifically, squeezed states of massive particles embody a key ingredient in tackling many of the above quests~\cite{Giovannetti_2004, Pirandola_2018,Degen_2017,Andersen_2016,Lawrie_2019, buonanno2004improving, Krisnanda2020observable}, and the development of simple approaches to generate such states becomes a pivotal step in the development of the field ~\cite{Abadie_2011,Aasi_2013,Pooser_2015,Samantaray_2017,Acernese_2019}. 

\begin{figure}[tb]
    \centering
    \includegraphics[width=0.95\linewidth]{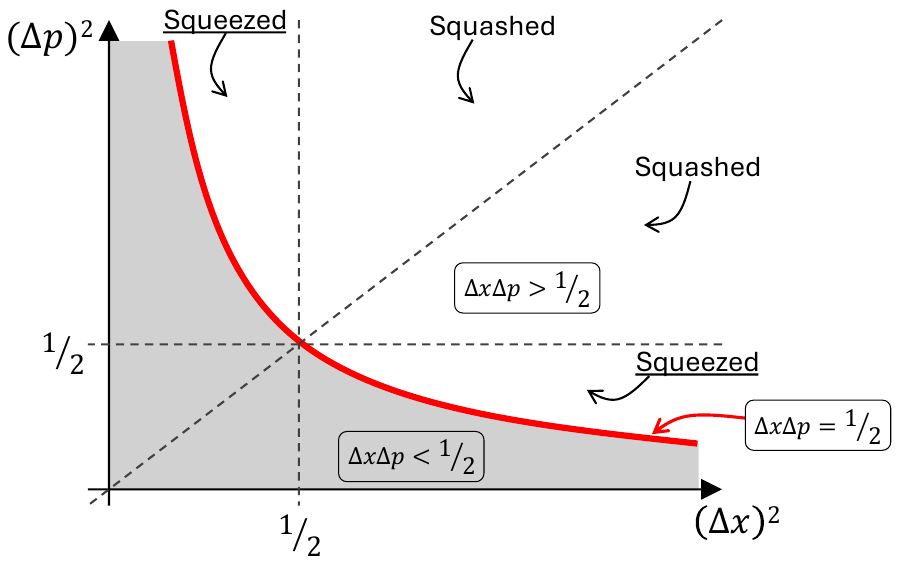}
    \caption{{
    Uncertainty space for the state of a single oscillator. For both squashed and squeezed states, the variance of a quadrature is smaller than that of its canonically conjugate. However, the smallest of the two variance of a squashed state is above the corresponding value for the ground state of the oscillator. For a squeezed state, such lower bound does no longer hold (the ultimate lower bound being set by the Heisenberg-Robertson uncertainty principle, shown by the thick red line in the figure).}
    }
    \label{fig:squashed_squeezed_states}
\end{figure}

Levitated nanomechanics offers a promising route to generate highly squeezed states of massive particles. In such a class of experiments, a nanoparticle is trapped within the waist of a focused laser beam, which provides a quadratic potential. 
Such systems have attracted a lot of interest in recent years: as the particle is levitated, interactions with the environmental phonons are suppressed, resulting in reduced damping and thermalisation rates.
Thus, the mechanical quality factor of the oscillator can reach values up to $10^{10}$ when operating in a high-vacuum chamber~\cite{Millen_2020,Gonzalez_2021}, allowing these systems to detect forces up to the attoNewton scale~\cite{Ranjit_2015}. This makes levitated mechanical systems an excellent platform for various quantum experiments, ranging from gravitational experiments that require high accuracies~\cite{Romero-Isart_2011,Bassi_2013,Balushi_2018,Ahn_2018}, to possible future generation of position superpositions with mesoscopic objects~\cite{muffato2024generation,Hornberger_2012,Scala_2013,Arndt_2014,Pino_2018}.
Squeezed states are a useful resource for all these experiments, as reduced position uncertainty enhances the signal-to-noise ratio for the detection, while states with increased position variance, i.e. large position superposition states, can be used in matter-wave interferometry~\cite{Millen_2020,Pino_2018}.
While the generation of squeezed states of light is routinely performed~\cite{wu1986generation, Andersen_2016}, that for massive levitated particles has proved to be a challenging task{, mostly due to the difficulty of preserving their quantum properties}.

For a refined study, here we distinguish squeezed states between ``squashed'' and genuinely squeezed states, {as illustrated in \cref{fig:squashed_squeezed_states}}. So far, only “squashed” states, \ie states whose smallest quadrature uncertainty is larger than that of the ground state, have been achieved in levitation experiments~\cite{Rashid_2016,Setter_2019}. Such class of states is fundamentally distinct from that of genuinely squeezed ones, where the reduction of uncertainty in one of the quadratures is below the zero-point fluctuation. 
Genuine quantum advantage only stems from genuinely squeezed states, which makes the conditions necessary to generate a truly squeezed state in a levitation experiment crucial for the purpose of demonstrating quantum advantages.

\begin{figure*}[tb]
    \subfloat{\label{fig:squeeze_protocol:a}}
    \subfloat{\label{fig:squeeze_protocol:b}}
    \centering
    \stackon[-0.5cm]{\includegraphics[width=\textwidth]{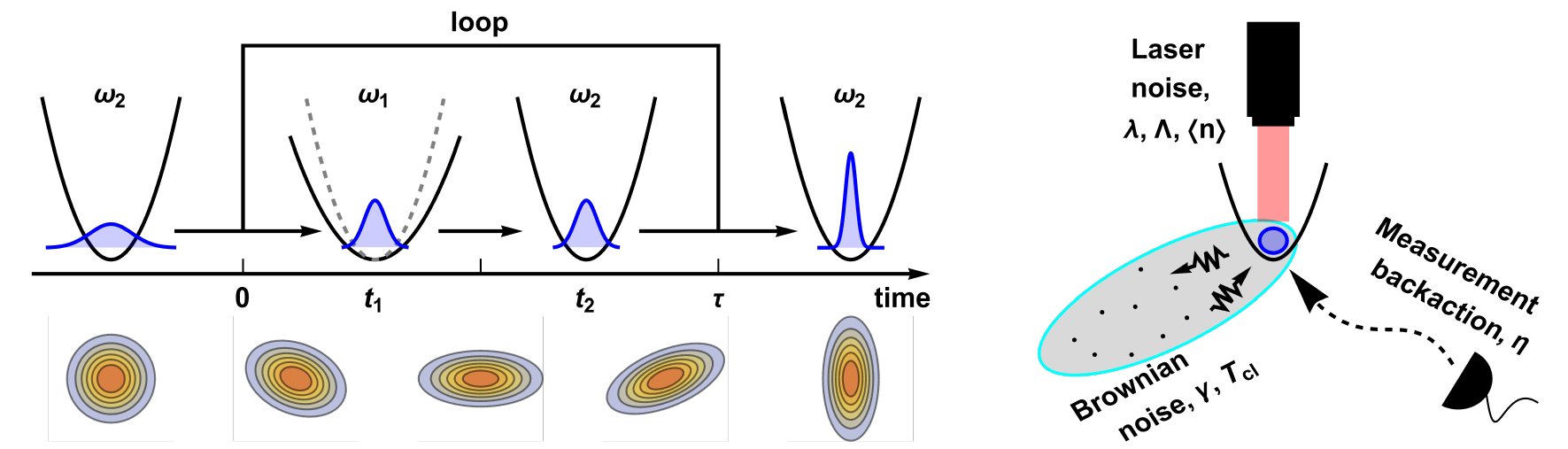}}{\textbf{a)}\hspace{11.2cm}\textbf{b)}\hspace{5.4cm}}
    \caption{Illustration of the squeezing protocol and the proposed setting. 
    \panel{a}: Squeezing protocol via frequency jump between $\omega_1$ and $\omega_2$ for a quantum harmonic oscillator (QHO). The time for a single squeezing cycle is $\tau=t_1+t_2$, while each $\omega_j$ is maintained for the time interval $t_j~(j=1,2)$. The full process might require multiple cycles to obtain the desired degree of squeezing.
    \panel{b}: Illustration of the potential sources of noises to the system. The levitated particle interacts with spurious optical modes, undergoes scattering from the residual gas in the vacuum chamber, and measurement backaction induced by the continuous monitoring system. The corresponding mechanisms are formally described by the dissipators in \cref{equ:all_dissipators}.
    }
    \label{fig:squeeze_protocol}
\end{figure*}

In this paper, we propose a protocol for the generation of highly squeezed motional states of a massive levitated particle. Our scheme leverages the dynamical switching between two frequencies of a quantum oscillator~\cite{Janszky_1986,Graham_1987,Janszky_1992} that has been employed to experimentally generate squeezed states of atomic systems~\cite{Xin_2021}. Time-modulation of trapping potentials was also used to generate high levels of squeezing between two position-position coupled oscillators~\cite{Cosco21} and trapped ions~\cite{Serafini2009}. We apply our protocol to the case of a continuously monitored levitated nanoparticle exposed to collisional, thermal and photon-recoil noises. We show that high levels of squeezing are achievable within a range of parameters compatible with current state-of-the-art setups. We also demonstrate the role of continuous monitoring in achieving squashed states and its apparent immateriality for the task of generating genuinely squeezed ones.
Our analysis also allows to establish the conditions that need to be achieved in order to quench the disrupting effects of the open dynamics.

The remainder of the paper is structured as follows. In \cref{sec:unitaryprotocol} we introduce the protocol for dynamical squeezing, including the implications of a continuous-measurement approach. The solution of the dynamics is given in \cref{sec:solution_of_the_system}, where we discuss the role that each of the experimental parameters has for the protocol efficiency. Finally, in \cref{sec:results_with_experiemental_settings} we discuss the experimental feasibility of our protocol and its potential implementation in realistic experimental settings.
\cref{sec:final} offers our concluding remarks and future perspectives.

\section{The squeezing protocol in open system dynamics}\label{sec:unitaryprotocol}

\noindent
\emph{The squeezing protocol -- }
The energy of the centre-of-mass motion of a nanoparticle trapped in a time-dependent quadratic optical potential is 
\begin{equation}\label{equ:hamiltonian}
    \hH_S(t) = \frac{\hp^2}{2m} + \frac{1}{2}m\omega^2(t) \hx^2,
\end{equation}
where $\hx$ and $\hp$ are the position and momentum operators of the quantum harmonic oscillator (QHO), $m$ is its mass, and $\omega(t)$ the trap frequency.
The squeezing protocol is performed by switching between two frequencies, $\omega_1$ and $\omega_2$, each being kept constant for a time interval $t_1$ and $t_2$ respectively~\cite{Janszky_1986}. Assuming $\omega_1<\omega_2$, the control protocol reads
\begin{equation}\label{equ:system_frequency}
    \omega(t) =
    \begin{cases}
        \omega_1, \quad 0+n\tau\leq t<t_1+n\tau,\\
        \omega_2, \quad t_1+n\tau\leq t <t_1+t_2+n\tau,
    \end{cases}    
\end{equation}
with $t_j = \pi/2\omega_j~(j=1,2)$, $\tau=t_1+t_2$ and $n$ labelling the number of squeezing cycles. We assume the system is initially prepared in the quadratic potential with $\omega(t) = \omega_2$. The effect of the squeezing protocol is illustrated in \cref{fig:squeeze_protocol:a} and detailed in \cref{apd:squeezing_protocol_unitary}. In particular, the squeezing amplitude after $N$ cycles is approximately $r = N\ln(\omega_1/\omega_2)/2$. Needless to say, such growth will not continue indefinitely, and the system variance will eventually stabilise due to the decoherence processes that we now address.

\noindent
\emph{Open dynamics under continuous measurement -- }
In a levitated nanoparticle experiment, the system is optically trapped by a laser and placed in a cold vacuum chamber. 
The levitated particle interacts with both the laser and the residual gas of the vacuum chamber, resulting in an open system dynamics, which therefore impacts the performance of the squeezing protocol.
In particular, we consider the following dissipators [cf. \cref{fig:squeeze_protocol:b}]
\begin{widetext}
\begin{subequations} \label[equation]{equ:all_dissipators}
    \begin{align}
    D_\text{cl}[\hr] &= -\frac{i\gamma}{2\hbar}[\hx,\{\hp,\hr\}] - \frac{\gamma m k_B T_\text{cl}}{\hbar^2} [\hx,[\hx,\hr]] 
     - \frac{\gamma}{16 m k_B T_\text{cl}}[\hp,[\hp,\hr]], \label{equ:caldeira_leggett_dissipator}\\
    D_\text{th}[\hr] &= \frac{i\lambda}{4\hbar} \left([\hp,\{\hx,\hr\}] - [\hx,\{\hp,\hr\}] \right)
    -\frac{\lambda(2\overline{n}+1)}{4\hbar m \omega} \left(m^2\omega^2[\hx,[\hx,\hr]]  + [\hp,[\hp,\hr]] \right), \label{equ:thermal_dissipator}\\[2ex]
    D_\text{lc}[\hr] &= -\Lambda [\hx,[\hx,\hr]], \label{equ:local_dissipator}
\end{align}
\end{subequations}
\end{widetext}
where $D_\text{cl}$ is the modified Caldeira-Leggett dissipator for the collisional noise that arises from the interaction between the system and surrounding residual gas \cite{breuer2002theory,Chang_2010}, $D_\text{th}$ is the thermalisation dissipator that arises from the interaction between the system and the optical modes from the laser, and $D_\text{lc}$ describes the decoherence in position due to photon recoils. Here $\gamma, \lambda, \Lambda>0$ are the respective coupling strengths, $m$ is the mass of the nanoparticle, $T_\text{cl}$ is the temperature of the chamber, and $\overline{n}$ is the mean excitation number of the particle. 

We assume that the position of the nanoparticle is continuously monitored. {Experimentally, this can be achieved through homodyne detection, \ie the collection of back-scattered photons that produces the photocurrent} \cite{Wiseman_Milburn_2009,Tebbenjohanns_2019,Magrini_2021,ALBARELLI2024129260}
\begin{equation}
    I(t)\df t = \sqrt{4\eta\Lambda}\braket{\hat{x}}\df t+\df W.
\end{equation}
{This quantity is proportional to the mean position of the system, shifted by the stochastic term $\df W$. We assume the latter to be a Gaussian random variable (corresponding to white noise), and model it as a Wiener increment}.
{The process of acquiring position information causes a backaction effect to the system that is accounted for by the innovation term $\sqrt{2 \eta \Lambda} H_{\hx}[\hr]\df W$, where $H_{\hx}[\hr] =\{\hx,\hr\} - 2\tr{\hx\hr}\hr$ describes the effect of the continuous position measurement, and $\eta$ is the measurement efficiency.
By combining the conditional dynamics with the dissipators as addressed above, the full stochastic master equation reads}
\begin{equation}\label{equ:full_master_equation}
     {\df}\hr = - \frac{i}{\hbar}[\hH_S,\hr]{\df t} + \sum_\nu D_\nu[\hr]{\df t} + \sqrt{2 \eta \Lambda} H_{\hx}[\hr]\df W,
\end{equation}
where the label $\nu =\text{cl},\text{th},\text{lc}$ refers to the superoperators in~\cref{equ:all_dissipators}. 

\noindent
\emph{Description of the dynamics --}
The process described in \cref{equ:full_master_equation} preserves the Gaussian nature of the input state throughout the time evolution. The evolved state can thus be fully characterised by first and second moments of the quadratures $\hat{x}$ and $\hat{p}$ of the system.
We define the mean vector $\bm{r} = (\braket{\hx},\braket{\hp})^T$ and the covariance matrix (CM) $\bm{\sigma}$ such that $\sigma_{\bm{r}_k, \bm{r}_j}=\frac{1}{2}\left(\braket{\hat{\bm{r}}_k\hat{\bm{r}}_j}+\braket{\hat{\bm{r}}_j\hat{\bm{r}}_k}\right)-\braket{\hat{\bm{r}}_k}\braket{\hat{\bm{r}}_j}$. 
{The dynamics of the first moments is stochastic, and depends on the measured photocurrent $I(t)$. However, such moments can be displaced to the origin of the phase space through a linear feedback control, a strategy that will be implicitly assumed henceforth. 
On the other hand, the CM of the conditional system satisfies the quantum Riccati equation} \cite{Genoni_2016} 
\begin{equation}\label{equ:riccati_equation}
    \dot{\bm{\sigma}} = \bm{A}\bm{\sigma} + \bm{\sigma} \bm{A}^T + \bm{D} - \bm{\sigma} \bm{B} \bm{B}^T\bm{\sigma},
\end{equation}
where we have introduced the drift, diffusion, and backaction matrices 
\begin{equation}\label{equ:riccati_equation_settings}
    \begin{aligned}
    \bm{A}=\begin{pmatrix}
      -a_1 & 1/m \\
      -m\omega^2 & -a_2
    \end{pmatrix}
    ,~
    \bm{D}=\begin{pmatrix}
      d_1 & 0 \\
      0 & d_2
    \end{pmatrix}
    ,~
    \bm{B}=\begin{pmatrix}
      0 & b \\
      0 & 0
    \end{pmatrix},
    \end{aligned}
    \end{equation}
    with
    \begin{subequations}
    \label[equation]{equ:riccati_equation_settings_matrices}
    \begin{align}
        a_1 &= \frac{1}{2}\lambda, \quad a_2=a_1 + \gamma,\quad b = \sqrt{\frac{8\eta\Lambda}{2\bar{n}+1}}, \\
        d_1 &=\frac{\hbar^2\gamma}{8 k_B m T_\text{cl}} + \frac{\hbar\lambda}{2 m\omega} \left(2\overline{n}+1\right), \\
        d_2 &=2\gamma k_B m T_\text{cl} + \frac{1}{2}\lambda\hbar m\omega\left(2\overline{n}+1\right) + 2\hbar^2\Lambda.
    \end{align}
\end{subequations}
In the drift matrix $\bm{A}$, the off-diagonal terms characterise the QHO system in \cref{equ:hamiltonian} given the mass and the trap frequency, while the diagonal terms characterise the damping rates in the mean position and momentum of the system. Characterising the drifting matrix $\bm{A}$, we find the time periods in \cref{equ:system_frequency} need to be modified for the open system dynamics, such that $t_j = \pi/ 2\Omega_j$ with $\Omega_j = \sqrt{\omega_j^2 - (a_1-a_2)^2/4}$. 
The diagonal terms of the diffusion matrix $\bm{D}$ are determined by the dissipation defined in \cref{equ:all_dissipators}. The continuous measurement leads to an additional term in the dynamics that is characterised by the matrix $\bm{B}$, which contains the efficiency $\eta$ of the continuous measurement. For $\eta =0$, \cref{equ:riccati_equation} reduces to the quantum Lyapunov equation.
The connection between the master equation and the Gaussian formalism is discussed in \cref{apd:sec:matrix_to_Gaussian_formalism}.

\section{Features of the asymptotic state to the protocol}\label{sec:solution_of_the_system}

Before investigating the performance of the squeezing protocol in a realistic experimental setting, we first study the asymptotic state of the squeezing protocol, whose dynamics are governed by \cref{equ:riccati_equation} for the cases with and without continuous measurement, respectively. 
In particular, we denote the position variances for the asymptotic CM without the squeezing protocol $\bm{\sigma}^{L (R)}_{\infty}$ as $\sigma^{L (R)}_{xx}$, that with the squeezing protocol $\bm{\sigma}^{L (R), \text{sq}}_{\infty}$ as $\sigma_{xx}^{L (R), \text{sq}}$. The superscript $L (R)$ stands for the case without (with) continuous measurement, {described by a Lyapunov (Riccati) equation for the CM.}

\noindent
\emph{Asymptotic state without continuous measurement -- }
Given the initial system CM $\bm{\sigma}_0$, the evolved system CM $\bm{\sigma}^L_t$ for $\eta=0$ is the solution of the quantum Lyapunov equation resulting from setting $\bm{B}=0$ in Eq.~\eqref{equ:riccati_equation}. This can be cast as 
\begin{subequations}\label[equation]{equ:langevin_equation_solution}
    \begin{gather}
        \bm{\sigma}^L_t = e^{t\bm{A}}(\bm{\sigma}_0-\bm{X}) e^{t\bm{A}^T}+\bm{X},
        \intertext{where the characteristic matrix $\bm{X}$ is the solution to the homogeneous equation}
        \bm{A} \bm{X} + \bm{X} \bm{A}^T + \bm{D} = \bm{0}.\label{equ:langevin_characteristic_matrix}
    \end{gather}
\end{subequations}
For a time-independent problem with constant drift and diffusion matrices $\bm{A}$ and $\bm{D}$, we have the asymptotic solution
\begin{equation}\label{equ:langevin_asymptotic_state}
    \bm{\sigma}^L_{t\to\infty} = \bm{X}.
\end{equation}
Under the action of the squeezing protocol in \cref{equ:system_frequency}, we are led to the time-evolved CM after $n$ squeezing loops
\begin{equation}\label{equ:langevin_squeezing}
    \bm{\sigma}^{L, \text{sq}}_{t = n\tau} = (\bm{S}^L_2\circ \bm{S}^L_1(\bm{\sigma}_0))^n,
\end{equation}
{with $\bm{S}^L_j(\bm{\sigma}) = e^{t_j\bm{A}_j}(\bm{\sigma}-\bm{X}_j) e^{t_j\bm{A}_j^T} + \bm{X}_j~(j=1,2)$ describing the squeezing operation when the trap frequency is set to $\omega_j$, matrices $\bm{A}_j$ and $\bm{X}_j$ the drift and characteristic matrices in \cref{equ:riccati_equation_settings,equ:langevin_characteristic_matrix}, and $\circ$ standing for the composition of operations}. 
Based on \cref{equ:langevin_squeezing}, the dynamics switches depending on the trap frequency, and corresponding asymptotic state with the squeezing protocol $\bm{\sigma}^{L, \text{sq}}_\infty$ can be computed numerically. 

\noindent
\emph{Asymptotic state with continuous measurement -- }
The case with continuous measurement is studied in a similar manner as that without it. Given the initial
system variance $\bm{\sigma}_0$, the evolved system variance $\bm{\sigma}^R_t$ following the quantum Riccati equation in \cref{equ:riccati_equation} can be solved analytically. The solution $\bm{\sigma}^R_t$ can be obtained via \cite{Ntogramatzidis_2011}
\begin{equation}\label[equation]{equ:riccati_equation_solution}
    \bm{\delta}_t - \bm{{\cal X}}_2 = e^{-t \bm{{\cal A}}^T}( \bm{\delta}_0 - \bm{{\cal X}}_2 ) e^{-t\bm{{\cal A}}},
\end{equation}
{where we define $\bm{\delta}_t = (\bm{\sigma}^R_t - \bm{{\cal X}}_1)^{-1}$, $\bm{{\cal A}} = \bm{A} - \bm{{\cal X}}_1 \bm{B} \bm{B}^T$. The characteristic matrices $\bm{{\cal X}}_1$ and $\bm{{\cal X}}_2$ satisfy the conditions}
\begin{equation}
\begin{aligned}
        &\bm{A} \bm{{\cal X}}_1 + \bm{{\cal X}}_1 \bm{A}^T + \bm{D} - \bm{{\cal X}}_1 \bm{B} \bm{B}^T\bm{{\cal X}}_1 = \bm{0}, \\
        &\bm{{\cal A}}^T\bm{{\cal X}}_2 + \bm{{\cal X}}_2 \bm{{\cal A}} - \bm{B}\bm{B}^T = \bm{0}. 
    \end{aligned}
\end{equation}
The derivation of such solution is shown in \cref{apd:sec:diffusion_equation_transform}.
For a time-independent QHO system with fixed matrices $\bm{A}$, $\bm{D}$ and $\bm{B}$, the system variance govern by \cref{equ:riccati_equation_solution} converges at $t\to\infty$, such that 
\begin{equation}\label{equ:riccati_asymptotic_state}
    \bm{\sigma}^{R}_{t\to\infty} = \bm{{\cal X}}_1 + \bm{{\cal X}}_2^{-1}.
\end{equation}
When the squeezing protocol is performed, the dynamics of the system variance is characterised by an expression similar to 
\cref{equ:langevin_squeezing}, namely
\begin{equation}\label{equ:riccati_squeezing}
    \bm{\sigma}^{R, \text{sq}}_{t = n\tau} = (\bm{S}^R_2\circ \bm{S}^R_1(\bm{\sigma}_0))^n,
\end{equation}
where the squeezing process $\bm{S}^R_{i}$ is given by \cref{equ:riccati_equation_solution} for the respective frequency $\omega_{i}$. The corresponding asymptotic state (which we compute numerically) is denoted as $\bm{\sigma}^{R, \text{sq}}_\infty$.

\begin{figure}[tb]
    \subfloat{\label{fig:squeezing_paramters:a}}
    \subfloat{\label{fig:squeezing_paramters:b}}
    \centering
    \stackon[-5.3cm]{\includegraphics[width=0.8\linewidth]{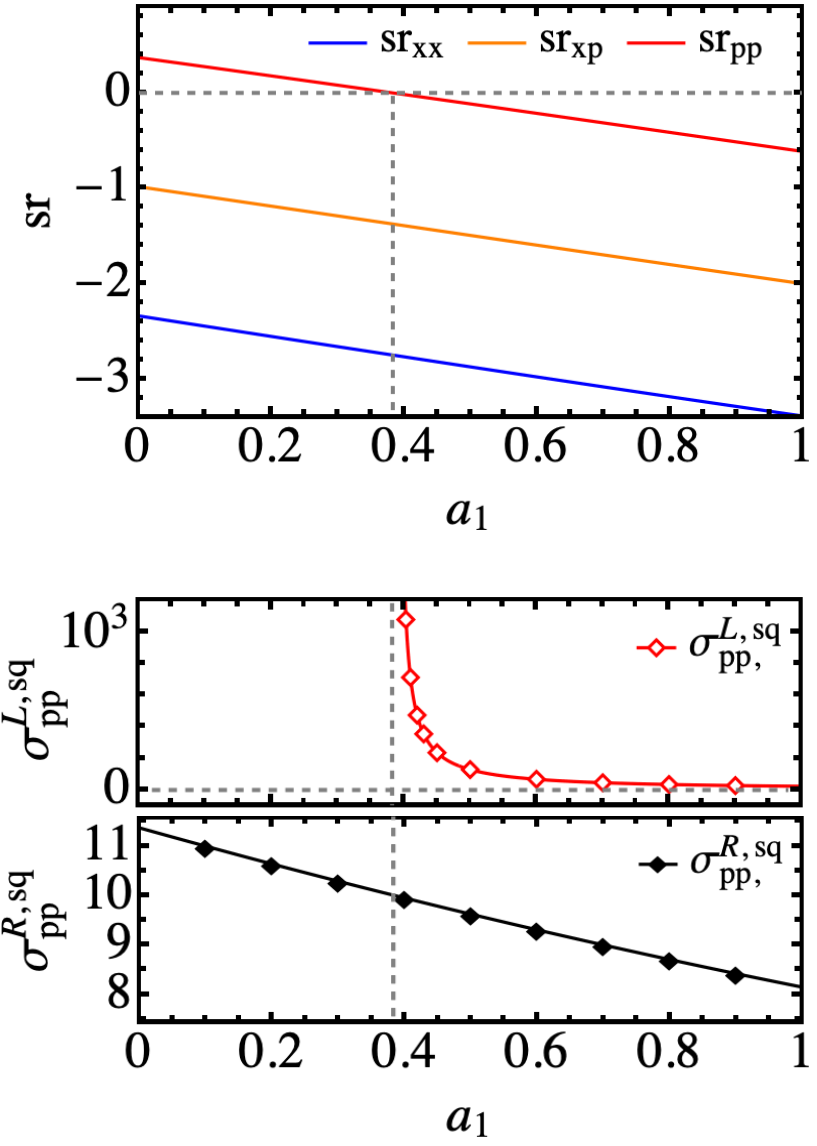}}{\stackon[-5.3cm]{\textbf{a)}}{\textbf{b)}}\hspace{7cm}}    
    \caption{
    Analysis of the asymptotic states. \panel{a} shows the squeezing parameters given by \cref{equ:momentum_squeezing_ratio} against the damping rate $a_1$. Here we set $2\omega_1=\omega_2=3\pi/2$, $m=a_2=1$ and $d_1=d_2=2$, from which we compute $a_1\approx 0.39$. \panel{b} shows the asymptotic momentum variance without continuous measurement (open diamonds, computed from \cref{equ:langevin_squeezing}) and with continuous measurement (filled diamonds, $b=2$, computed from \cref{equ:riccati_squeezing}).
    }
    \label{fig:squeezing_paramters}
\end{figure}

\begin{figure*}[tb]
    \subfloat{\label{fig:config_langevin_riccati:a}}
    \subfloat{\label{fig:config_langevin_riccati:b}}
    \subfloat{\label{fig:config_langevin_riccati:c}}
    \subfloat{\label{fig:config_langevin_riccati:d}}
    \centering
    \stackon[-0.54cm]{\includegraphics[width=\textwidth]{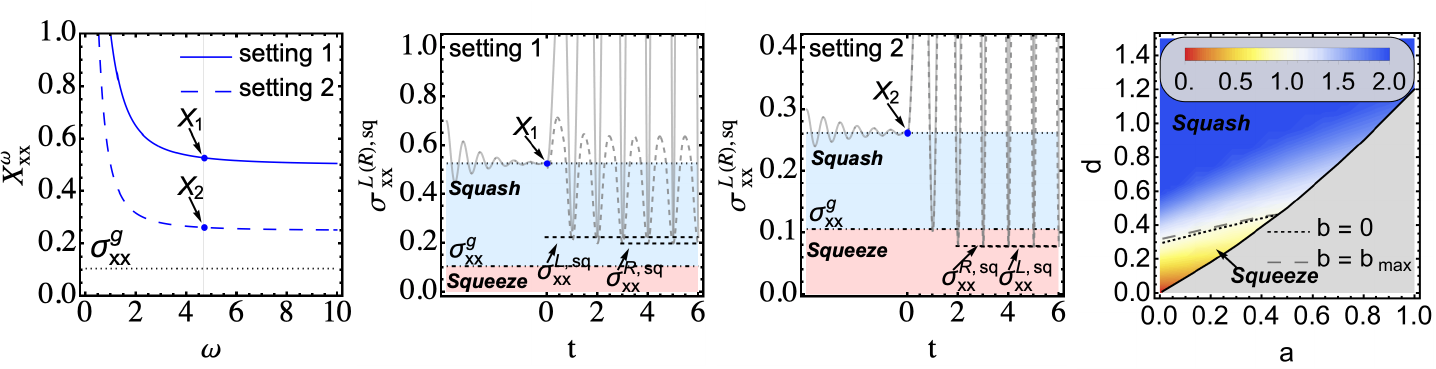}}{\textbf{a)}\hspace{4.18cm}\textbf{b)}\hspace{4.08cm}\textbf{c)}\hspace{4.22cm}\textbf{d)}\hspace{4.08cm}}
    \caption{{Investigation on the position variances for the asymptotic states and the capability of the squeezing protocol given different experimental settings. \panel{a} shows the asymptotic position variance before the squeezing protocol $X^\omega_{xx}$ in} \cref{equ:langevin_asymptotic_state} {against the trap frequency $\omega$ in} \cref{equ:hamiltonian}. 
    {Here the ground state variance at frequency $\omega_2$ is $\sigma^g_{xx} = 1/3\pi$. 
    Panels~\textbf{b)} and \textbf{c)} demonstrate the dynamics of the system position variance when the squeezing protocol is performed. Here the system starts from an arbitrary state and equilibrates to $X_1$ ($X_2$) at the time $t=0$. Then, the squeezing protocols without continuous measurement (solid gray line) and with continuous measurement (dashed gray line, $b=b_{max}$) are applied, where ${\sigma}^{L,\text{sq}}_{xx}$ and ${\sigma}^{R,\text{sq}}_{xx}$ are the corresponding minimum position variances. Here, we say that the system is squashed if $\sigma^g_{xx}<\sigma^{L(R), \text{sq}}_{xx}<X_1~(X_2)$ (shaded blue region), and that it is squeezed if $\sigma^{L(R),\text{sq}}_{xx}<\sigma^g_{xx}$ (shaded red region). 
    \panel{d} shows the  ratio between the variances for the squeezed asymptotic state and the ground state, \ie $\sigma^{L (R), \text{sq}}_{xx} / \sigma^g_{xx}$, given different values of $a$, $d$ with the gray region forbidden by} \cref{equ:gaussian_state_conditions}. {Here the blue region indicates the settings that only achieve squashing, while the red region those that achieve squeezing. The borders between two regions (\ie when $\sigma^{L (R), \text{sq}}_{xx} = \sigma^g_{xx}$) are indicated by the black dotted line (in the case without measurement, $b=0$) and gray dashed lines (in the case with measurement with $b=b_{\max}$). }
    }
    \label{fig:config_langevin_riccati}
\end{figure*}

\noindent
\emph{Features of the asymptotic states --}
The squeezing process described by \cref{equ:langevin_squeezing} may cause an unstable asymptotic state due to the expansion of one quadrature during the squeezing of the other quadrature. This issue can be circumvented by applying continuous measurement along with the squeezing protocol as the dynamics described by \cref{equ:riccati_squeezing}. Indeed, suppose the variance of a system's quadrature $\sigma_n$ follows the relation $\sigma_{n+1} = e^{sr}\sigma_n + \chi$ at $n$-th squeezing cycle, where ${sr}$ is the squeezing parameter and $\chi>0$ is the amount of diffusion over one squeezing cycle. The quadrature variance $\sigma_\infty$ approaches to a steady point ${\chi}/{(1-e^{sr})}$ if $sr<0$ and to the infinity if $sr\geq0$. Experimentally, the infinity variance means that the system can become unstable after a number of squeezing cycles, when one quadrature of the system expands too much to be well contained by the trap. 

We demonstrate this issue with the system's momentum variance, when the system is squeezed in its position. Without applying the continuous measurement, \cref{equ:langevin_squeezing} gives the squeezing parameters for each term in the CM, where for the momentum spread one has 
\begin{equation}\label{equ:momentum_squeezing_ratio}
    {sr}_{pp} \approx - \frac{\pi(a_1+a_2)(\Omega_1+\Omega_2)}{2\Omega_1\Omega_2} + \ln\left(\frac{\Omega_2^2}{\Omega_1^2} + O(\Delta a) \right),
\end{equation}
with $\Delta a = a_2-a_1>0$ and $\Omega_i=\sqrt{\omega_i^2-(a_1-a_2)^2/4}$. The full expressions for all squeezing parameters are given in \cref{apd:sec:squeezing_paramters}. Given the definition $\omega_1<\omega_2$ (hence $\Omega_1<\Omega_2$), and the small damping regime $\Delta a\to 0$, we notice it is possible to have ${sr}_{pp}>0$ when the damping rate $a_1$ is small, while the other two parameters ${sr}_{xx}$ and ${sr}_{xp}$ remain negative for any setting of $\Omega_{1,2}$ and $a_{1,2}$, as illustrated by \cref{fig:squeezing_paramters:a}.
This leads to the explosion of the system momentum variance (whose asymptotic value goes to infinity) in the regime of high quality factors $Q \sim \omega_2/a_1$. 
This issue can be circumvented when the continuous measurement is applied to the system, as shown by the open and filled diamonds in \cref{fig:squeezing_paramters:b}, representing the cases without and with continuous measurement ($b=2$), respectively. 
Therefore, we conclude that continuous measurement is crucial for applying the squeezing protocol in the high-quality regime, since it guarantees that the system under the squeezing protocol will always reach a stable state with a finite position and momentum variances.

\noindent
{\emph{Features of the squeezing protocol --}
Here we explain our squeezing protocol using a toy model. To keep the outcomes meaningful, we examine the state of the system at any time with the conditions \cite{WANG_20071, Adesso_2014} 
\begin{equation}\label{equ:gaussian_state_conditions}
    \bm{\sigma}\geq0, \quad \bm{\sigma} + i \bm{\Omega}\geq 0, \quad \text{and}\quad \left(2\sqrt{\text{Det}[\bm{\sigma}]}\right)^{-1}\in[0,1].
\end{equation}
The first corresponds to the positivity of the state, the second is the verification of the Heisenberg-Robertson uncertainty principle, while the last measures the purity of the state of the system, where the symplectic matrix reads $\bm{\Omega}={\begin{psmallmatrix}0 & 1\\-1 & 0\end{psmallmatrix}}$. We take dimensionless units $\hbar=m=1$ with setting~1:~$a_{1,2}=0.5$, $d_{1,2}=1$; setting~2: $a_{1,2}=d_{1,2}=0.3$, and choose $b=b_{\max}$ (the maximum value allowed by} \cref{equ:gaussian_state_conditions}) {for the matrices $\bm{A}$, $\bm{D}$ and $\bm{B}$ given by} \cref{equ:riccati_equation_settings_matrices}. {We take two oscillating frequencies to be $2\omega_1=\omega_2=3\pi/2$ and show all the results in \cref{fig:config_langevin_riccati}. }

{In the preparation stage, we assume the system is not continuously measured. The system is initially stabilised at a fixed trap frequency $\omega$ [i.e.~the frequency appearing in} \cref{equ:hamiltonian}], {whose position variance $X^{\omega}_{xx}$ is shown by} \cref{fig:config_langevin_riccati:a}. {In our toy model, the state of the system is initially stabilised at frequency $\omega_2$ with the initial position variance $X_{1,2}$ for settings $1,2$ (blue dots) and the corresponding ground state position variance $\sigma^g_{xx}={\hbar}/{(2m\omega_2)}= 1/3\pi$ (dot-dashed black line). 
Clearly, the position variance for the stabilised state $X_1$ ($X_2$) is far above that of the ground state $\sigma^g_{xx}$. }

{The performance of the squeezing protocol addressed in this paper is demonstrated in} \cref{fig:config_langevin_riccati:b,fig:config_langevin_riccati:c}, {where we oscillate between the trap frequencies $\omega_1$ and $\omega_2$ and the time for one squeezing loop $t_1+t_2=1$. Starting from an arbitrary state, the system is initially stabilised (no measurement) at the fixed frequency $\omega_2$ with the position variance $X_1$ ($X_2$) at $t<0$. Then, we perform the squeezing protocol starting at $t=0$, with and without the continuous measurement. 
Here, the continuous gray line represents the case without continuous measurement, and the gray dashed line represents the case with continuous measurement ($b=b_{max}$ allowed by} \cref{equ:gaussian_state_conditions}). {The minimal position variance without continuous measurement is labelled by $\sigma^{L, \text{sq}}_{xx}$, that with continuous measurement is labelled by $\sigma^{R, \text{sq}}_{xx}$, and the ground-state variance is given by $\sigma^g_{xx}$.
In setting 1 ($b_{\max} = 1.9$,} \cref{fig:config_langevin_riccati:b}), {we only achieve quantum squashing for both cases, \ie $\sigma^g_{xx}<\sigma_{xx}^{R, \text{sq}}<\sigma_{xx}^{L, \text{sq}}$, due to large damping and diffusion terms $a_{1,2}$ and $d_{1,2}$.  
In setting 2 ($b_{\max}=0.7$,} \cref{fig:config_langevin_riccati:c}) {we see that quantum squeezing can be achieved in both cases.
We can observe the effects of the continuous measurement here. Other than the improvement to the squashing $\sigma_{xx}^{R, \text{sq}}<\sigma_{xx}^{L, \text{sq}}$, we find the overall amplitude of the oscillation of the system is reduced when continuous measurement is performed. This is due to the noise reduction from the continuous measurement, which is characterised by 
a larger purity in the case  under continuous measurement. 
This could be advantageous as it suggests that in this case the system remains localised in a smaller region.}

{Stringent conditions are required to achieve a truly quantum squeezing, and such conditions for our toy model is shown in} \cref{fig:config_langevin_riccati:c}. {Here the colour represents the ratio between the position variances for the asymptotic state without continuous measurement and the ground state, \ie $\sigma^{L, \text{sq}}_{xx} / \sigma^g_{xx}$, under different settings (we take $a_1=a_2=a$ and $d_1=d_2=d$ constrained by} \cref{equ:gaussian_state_conditions}). {For the values of the parameters in the blue region, the protocol can only squash the system. Conversely, for those in the red region, the protocol can squeeze the system below the ground state variance. The border between the blue and red regions is indicated by the black dotted line, on which one has $\sigma^{L, \text{sq}}_{xx} = \sigma^g_{xx}$. 
The continuous measurement can slightly improve the squeezing, as indicated by the border of $\sigma_{xx}^{R,\text{sq}}=\sigma^g_{xx}$ (dashed gray lines, $b_{\max}=1.9$) moving towards the blue region.
}

\section{Testing in experimental settings}\label{sec:results_with_experiemental_settings}

\renewcommand{\arraystretch}{1.3}
\begin{table}[b]
    \centering
    \begin{tabular}{|c|c|c c|c|c|c|c|c|c|}
        \hline
        $m$ & $\omega/2\pi$ & $\gamma$ & $\lambda$ & $\Lambda$ & $T_\text{cl}$ & $\overline{n}$ & $\eta$  \\
        (\si{\fg}) & (\si{\kHz}) & (\si{\Hz}) & (\si{\Hz}) & (\si{\m^{-2}\Hz}) & (\si{\K}) & &\\
        \hline
        1 & $50 \sim 100$ & \multicolumn{2}{c|}{$\geq 10^{-7}$} & $10^{23}\sim 10^{26}$ & $50$ & $10^7$ & $\leq0.3$ \\
        \hline
    \end{tabular}
    \caption{Collection of the experimental parameters and their values for the discussion in \cref{sec:results_with_experiemental_settings}. The connection of the parameter to the respective environmental noise is illustrated in \cref{fig:squeeze_protocol:b}.}
    \label{tab:parameter_values}
\end{table}

To infer the capability of our squeezing protocol in a real experiment, we substitute the parameters in \cref{equ:riccati_equation_settings} with those that can be found in recent experiments \cite{Rashid_2016,Setter_2019,Magrini_2021,Tebbenjohanns_2021,Militaru_2022,Magrini_2022,Dania_2023}. As a reference, we consider the following setting.
Suppose a silica nanoparticle of radius $R = 50$\,\si{\nm} and mass $1$\,\si{\fg} ($\rho = 2200$\,\si{\kg/\m^3)} is levitated in an optical trap, which can oscillate between two frequencies $\omega_1/2\pi = 50$\,\si{\kHz} and $\omega_2/2\pi = 100$\,\si{\kHz}. The experiment can be performed in a cryostat and ultra-high-vacuum environment, such that the environment temperature $T_\text{cl} = 50$\,\si{\K}, and the quality factor $Q = \omega/\gamma$ can be as high as $10^{10}$ \cite{Dania_2023}. Thus, the damping rates (from the collisional and thermal noises) are weak and {they can be estimated by $\gamma\sim\lambda\sim 10^3(P/\si{\mbar})\,\si{\Hz}$}, where we assume that the chamber pressure can go as low as $P\geq 10^{-10}$\,\si{\mbar} \cite{Dania_2023}. The photon-recoil rate can be estimated by the equation \cite{Gardiner_2004,Seberson_2020,Gonzalez-Ballestero_2019}
\begin{equation}\label{equ:photon_recoil_rate}
    \Lambda=\frac{7 \pi \varepsilon_0}{30\hbar} \left(\frac{\epsilon_c V E_t}{2\pi}\right)^2 k_0^5,
\end{equation}
where $\varepsilon_0$ is the vacuum permittivity, $\epsilon_c=3{(\varepsilon-1)}/{(\varepsilon+2)}$ is written in terms of the the relative dielectric constant $\varepsilon$ of the nanoparticle, whose volume is $V$, $k_0=\omega_0/c$ with $c$ the speed of light and $\omega_0 = {2\pi c}/{\lambda}$ the laser beam frequency.
Following the analysis in \cref{apd:sec:range_of_coefficients}, we estimate $\Lambda \sim 10^{26}$\,\si{\m^{-2}\Hz}~\cite{Malitson_1965,Gonzalez-Ballestero_2019,Windey_2019}. {We take the mean occupation number of the particle at temperature $50$\,\si{K} to be $\overline{n}=10^7$. However, the occupation number can be further reduced to $\sim0.5$ with specific cooling techniques \cite{Delic_2020}, which further reduce the effect of the thermal noise.} We estimate that the efficiency of the measurement is no more than $30\%$ \cite{Dania_2022}, and summarise these values in \cref{tab:parameter_values}.

\begin{figure}[tb]
    \subfloat{\label{fig:experimental_squeezing:a}}
    \subfloat{\label{fig:experimental_squeezing:b}}
    \subfloat{\label{fig:experimental_squeezing:c}}
    \centering
    \stackon[-9.88cm]{\includegraphics[width=0.85\linewidth]{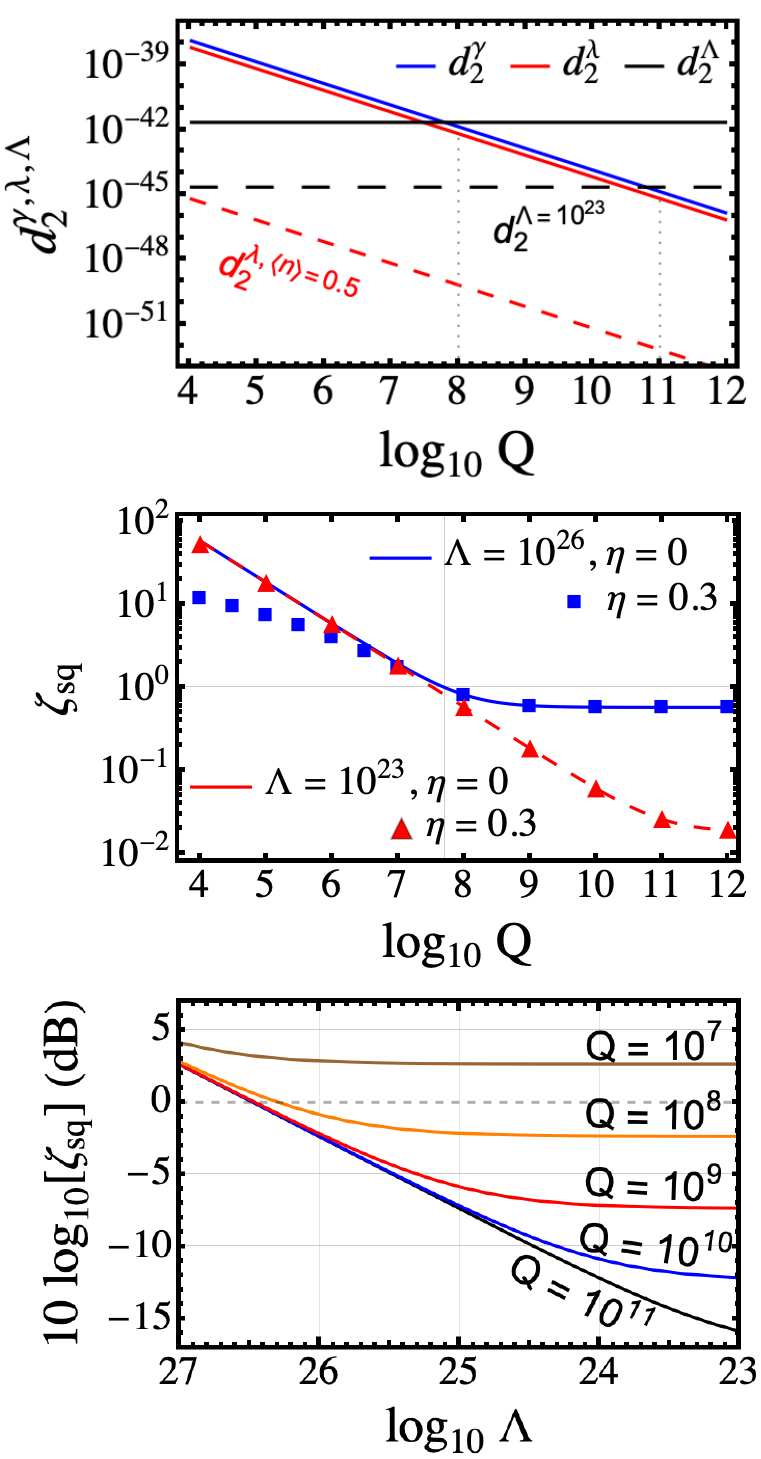}}{
    \stackon[-9.8cm]{\textbf{a)}}
    {\stackon[-4.95cm]{\textbf{b)}}{\textbf{c)}}}
    \hspace{7cm}
    }    
    \caption{Contributions of the noises and the squeezing ratio. \panel{a} shows the quality factor $Q=\omega/\gamma (\lambda)$ dependence of the noise contributions from \cref{equ:d2_components}, which rise from the collisional noise ($d_2^\gamma$), the thermal noise ($d_2^\lambda$) and the photon-recoil noise ($d_2^\Lambda$). 
    \panel{b} shows the logarithmic relation of the squeezing ratio $\zeta_\text{sq} =\sqrt{\sigma_{xx}^\text{sq}(\tau)/\sigma_{xx}^g}$ against the quality factor $Q$. The blue (red dashed) line represents the cases for $\Lambda = 10^{26}\, (10^{23})$\,\si{\m^{-2}\Hz} with $\eta = 0$, and the square (triangle) represents the corresponding case with $\eta = 0.3$. 
    {\panel{c} shows the amount of squeezing $10 \log_{10} (\zeta_\text{sq})=10 \log_{10} (\sqrt{\sigma_{xx}^\text{sq}(\tau)/\sigma_{xx}^g})$ in decibel against the photon-recoil noise rate $\Lambda$ for different quality factors $Q$.} 
    }
    \label{fig:experimental_squeezing}
\end{figure}

We first investigate the contributions of the noises given the setting discussed above. In particular, we scrutinize $d_2$ in \cref{equ:riccati_equation_settings} and compare the order of magnitudes for each terms. We find the following values (in unit \si{\m^2\kg^2\Hz^3})
\begin{subequations}\label[equation]{equ:d2_components}
    \begin{align}
        d_2^\gamma &=2\gamma k_B m T_\text{cl} \sim 10^{-45},\\
        d_2^\lambda &=\overline{n}\lambda\hbar m\omega\sim 10^{-46},\\
        d_2^\Lambda &=2\hbar^2\Lambda \sim 10^{-42},
    \end{align}
\end{subequations}
where we have taken $\gamma=\lambda=10^{-6}$\,\si{\Hz}, $\Lambda=10^{26}$\,\si{\m^{-2}\Hz}, and $\overline{n}=10^7$. This suggests that the strength of collisional and thermal noises are comparable at $T_\text{cl}=50$\,\si{\K}, and they can be smaller than the photon-recoil noise for large enough qualify factor $Q$. Indeed, collisional, thermal and photon-recoil noises are comparable when $Q\sim 10^8$, as shown by \cref{fig:experimental_squeezing:a}. 
On the other hand, if the photon-recoil rate is reduced by $3$ orders, \ie $\Lambda \sim 10^{23}$\,\si{\m^{-2}\Hz}, collisional and photon-recoil noises are comparable when $Q\sim 10^{11}$.

We then consider the following squeezing experiment [cf.~\cref{fig:config_langevin_riccati:b}]. The levitated particle firstly equilibrates at trap frequency $\omega_2$, such that the centre of motion of the particle is a QHO system in the asymptotic state given by \cref{equ:langevin_asymptotic_state}. Then, the squeezing protocol is applied to the system at $t=0$, and we measure its position variance at the end of the protocol when $t = \tau$. We denote the position variance of the system at time $t$ to be $\sigma^\text{sq}_{xx}(t)$, and take the position variance of the ground state $\sigma^\text{g}_{xx}$ at frequency $\omega_2$ as the reference. In the best scenario (corresponding to $\gamma=\lambda=10^{-6}$\,\si{\Hz}, $\Lambda=10^{23}$\,\si{\m^{-2}\Hz}, $\overline{n}=10^7$), we have for the initial system
\begin{subequations}
    \begin{align}
        \sigma^\text{sq}_{xx}(0) = X_{xx}^{\omega_2} & \approx 3.1\times10^{3}\,\si{\nm^2}, \\
        \sigma^\text{g}_{xx}=\frac{\hbar}{2 m \omega_2}& \approx 8.4\times10^{-5}\,\si{\nm^2}.
    \end{align}
\end{subequations}
Hence,  the system has an initial position spread being a factor $\zeta_\text{sq}(0) = \sqrt{\sigma^\text{sq}_{xx}(0)/\sigma^\text{g}_{xx}} \sim 10^{4}$  larger than the ground state. In the worst scenario ($\Lambda=10^{26}$\,\si{\m^{-2}\Hz}), such a factor becomes $\zeta'_\text{sq}(0)\sim 10^{5}$. 
After applying the squeezing protocol, the degree of squeezing in a realistic experimental settings is demonstrated by the ratio $\zeta_\text{sq} =\sqrt{\sigma_{xx}^\text{sq}(\tau)/\sigma_{xx}^g}$ in \cref{fig:experimental_squeezing:b}, where a value of $\zeta_\text{sq}<1$ means the position variance of the system is squeezed below that of the ground state. We consider two cases for the the photon-recoil rate, \ie $\Lambda = 10^{26}$ and $10^{23}$\,\si{\m^{-2}\Hz}, {without continuous measurement ($\eta =0$)}. For small quality factors $Q$, we see that the ratio $\zeta_\text{sq}$ coincides in both cases (blue continuous and red dashed lines), since here the dynamics is dominated by the collisional and thermal noises. Conversely, we see a plateau (at $Q>10^8$ for $\Lambda = 10^{26}$\,\si{\m^{-2}\Hz}, and at $Q>10^{11}$ for $\Lambda = 10^{23}$\,\si{\m^{-2}\Hz}) when the photon recoils becomes the dominant noise contribution [\cf \cref{fig:experimental_squeezing:a}]. In both cases, the squeezing ratio $\zeta_\text{sq}$ at high quality factor is below 1, indicating that one can achieve genuine squeezing in a levitation experiment at ultrahigh vacuum. Based on our model, we have $\zeta_\text{sq} \approx 0.58$ at $Q=10^{12}$ for $\Lambda = 10^{26}$\,\si{\m^{-2}\Hz} and $\zeta_\text{sq} \approx 0.02$ at $Q=10^{12}$ for $\Lambda = 10^{23}$\,\si{\m^{-2}\Hz}.

Squares and triangles in \cref{fig:experimental_squeezing:b} represent the cases taking the continuous measurement with $\eta =0.3$, which only provide a small improvement to the squeezing (\eg $\zeta_\text{sq}^{\eta=0.3} \approx 0.57$ at $Q=10^{12}$ for $\Lambda = 10^{26}$\,\si{\m^{-2}\Hz}), which cannot be seen using the logarithmic scale. Indeed, the effect of the continuous measurement becomes  prominent only when the photon-recoil noise is large ($\Lambda=10^{26}$\,\si{\m^{-2}\Hz}) and the qualify factor is low ($Q<10^8$), thus enhancing the squashing effect. If the photon-recoil noise can be reduced (\eg when $\Lambda=10^{23}$\,\si{\m^{-2}\Hz}), the continuous measurement would not produce any improvement in the squeezing even for the low qualify factor [\cf red line and triangles in the figure]. Indeed, the continuous measurement effect described by \cref{equ:full_master_equation} is proportional to the photon-recoil noise $\Lambda$. 

{The achievable values for the state-of-art optical levitation experiment are $Q\sim 10^{10}$ and $\Lambda\sim10^{26}$\,\si{\m^{-2}\Hz}. \cref{fig:experimental_squeezing:c} shows the relation between the degree of squeezing and the photon-recoil noise around this region. We measure the squeezing in decibel (dB) by $10 \log_{10} (\zeta_\text{sq})$, such that squeezing below the ground-state variance gives negative values. 
We see that squeezing is only possible when $Q\geq 10^8$, and the maximal achievable squeezing is approximately $-2.5$\,\unit{\dB}. We can see that, at the current noise level ($\Lambda\sim10^{26}$\,\si{\m^{-2}\Hz}), improving the quality factor $Q$ from $10^9$ to $10^{10}$ only slightly improves the amount of squeezing. Therefore, it is more prominent to reduce the photon-recoil noise level in order to acquire more below-the-ground-state squeezing (\eg $10 \log_{10}(\zeta_\text{sq})\approx-7.35$\,\unit{\dB} at $\Lambda\sim10^{25}$\,\si{\m^{-2}\Hz} and $Q\sim10^{10}$).}

\section{Conclusions}\label{sec:final}
We have proposed a squeezing protocol (via frequency jumps) applied to a levitated nanoparticle subjected to continuous monitoring of its position. The dynamics of the system is influenced by the collisional, thermal, and photon-recoil noises, and influenced by the stochastic noise caused by the continuous measurement. We have estimated the potential for achieving large squeezing of the considered massive quantum system, considering parameters from recent experiments. We have found that while the photon-recoil noise plays a dominant role, in the high quality-factor regime, the position spread of the system can still be squeezed below the ground state variance. 
The backaction from continuous measurement does not help squeezing performances. Our study addresses the engineering of genuine quantum resources for sensing and metrology in levitated optomechanics, while also providing a route to the achievement of states that will be crucial for investigations on the foundations of quantum mechanics. The method illustrated here will benefit of the combination with control methods based on the modulation of the environmental properties as proposed in Ref.~\cite{delCampo2020}. Another possible direction of exploration goes along the line of embedding the oscillator in a continuously monitored optical cavity, as in Ref.~\cite{Genoni2015}. The closeness of our assessment to experimental reality paves the way to a ready implementation of the scheme.

\acknowledgments
We acknowledge support by the European Union's Horizon Europe EIC-Pathfinder project QuCoM (101046973), 
the Leverhulme Trust (grants RPG-2022-57 and RPG-2018-266), the Royal Society Wolfson Fellowship (RSWF/R3/183013), the UK EPSRC (grants EP/W007444/1, EP/V035975/1, EP/V000624/1, EP/X009491/1, EP/T028424/1), the Department for the Economy Northern Ireland under the US-Ireland R\&D Partnership Programme, and the PNRR PE National Quantum Science and Technology Institute (PE0000023). We
further acknowledge support from the QuantERA grant
LEMAQUME, funded by the QuantERA II ERA-NET
Cofund in Quantum Technologies implemented within
the EU Horizon 2020 Programme

\bibliography{references}

\appendix

\section{The squeezing protocol}\label{apd:squeezing_protocol_unitary}

Squeezing of the motion of a harmonic oscillator through the dynamical switching of frequency has been proposed in Ref.~\cite{Janszky_1986,Graham_1987,Janszky_1992}, and has been employed to experimentally generate squeezing in atomic systems~\cite{Xin_2021}. This operation is based on the Bogoliubov transformation which describes the effect of switching the frequency of the oscillator.
Given the Hamiltonian in \cref{equ:hamiltonian}, we call $\{\ha,\ha^\dag\}$ ($\{\hb,\hb^\dag\}$) the ladder operators corresponding to frequency $\omega_1$ ($\omega_2$).
They satisfy the relations
\begin{align}
    \hb=\mu \ha + \nu \ha^\dag, \qquad
    \hb^\dag=\mu \ha^\dag + \nu \ha,
\end{align}
where $\mu=\cosh(r)=(\omega_1+\omega_2)/{2\sqrt{\omega_1\omega_2}}$, $\nu = \sinh(r)=(\omega_1-\omega_2)/{2\sqrt{\omega_1\omega_2}}$ and $r=\frac{1}{2}\ln(\omega_2/\omega_1)$. The Bogoliubov transformation is linked to the squeezing operation such that $\hb=\hat{S}^\dag(r)\ha \hat{S}(r)$ with
\begin{equation}\label{apd:equ:squeezing_operator}
    \hat{S}(r) = \exp\left[\frac{1}{2}(r \ha^2- r\ha^{\dag 2})\right].
\end{equation}
Therefore, switching the frequency of the oscillator leads to a squeezing operation on the system.

To achieve squeezing in the position of the system, a squeezing protocol is defined by \cref{equ:system_frequency} in the main text, which gives the equations for the expectation values of the quadrature operators $\hx$ and $\hp$ as
\begin{subequations}\label[equation]{apd:equ:quadrature_unitary}
    \begin{align}
         \frac{\df}{\df t}\braket{\hx} &= \frac{\braket{\hp}}{m},\\
         \frac{\df}{\df t}\braket{\hp} &= -m\omega^2\braket{\hx}.
    \end{align}
\end{subequations}
Indeed, in each time period $t_j$, the dynamics of the quadratures in \cref{apd:equ:quadrature_unitary} can be solved analytically with $\omega=\omega_j$. By defining the initial mean position and  momentum of the system as $x_0 = \braket{\hx(0)}$ and $p_0=\braket{\hp(0)}$, the evolution results in the transformation $(x_0,p_0)^T \mapsto \bm{M}_j(t)(x_0,p_0)^T$ with
\begin{equation}
    \bm{M}_j(t)=\begin{pmatrix}
    \cos (\omega_j  t) &  \sin(\omega_j  t)/m\omega_j \\
    -m\omega_j\sin(\omega_j t) & \cos(\omega_j  t)
    \end{pmatrix}.
\end{equation}
This transformation leads to the combined action of  rotation and squeezing of the system. 
In particular, the protocol defined in \cref{equ:system_frequency} acts on the system in the following way [cf. \cref{fig:squeeze_protocol} where the protocol is illustrated]: when $\omega = \omega_1$, the system is squeezed along the $\hx$ quadrature. When $\omega=\omega_2$, the system is squeezed along the $\hp$ quadrature. However, after the first part of squeezing $t\geq t_1=\pi/2\omega_1$, the quadratures of the system swap due to the rotation of the system, $\hx\leftrightarrow \hp$. Therefore, the second part of the protocol acts on the same quadrature as the first part does. After time $\tau$, this gives the transformation of the average values of the quadratures $x_\tau = \braket{\hx(\tau)}$ and $p_\tau=\braket{\hp(\tau)}$ as
\begin{equation}\label[equation]{apd:equ:combined_action}
    \begin{aligned}
        (x_\tau,p_\tau)^T &= \bm{M}_2(t_2)\circ\bm{M}_1(t_1)~(x_0,p_0)^T \\
        &= -\left(e^{r} x_0, e^{-r} p_0\right),
    \end{aligned}
\end{equation}
where we have defined $e^{r}\equiv \omega_1/\omega_2$, and the squeezing parameter \(r=\ln{\omega_1}/{\omega_2}\). \cref{apd:equ:combined_action} results in squeezing of the position quadrature, so that the variance $\sigma_{xx}(0) = \braket{\hx^2(0)}$ transforms to
\begin{equation}\label{apd:equ:unitary_outcome}
    \sigma_{xx}(\tau)=e^{2r}\sigma_{xx}(0),
\end{equation}
where we set $\braket{\hx(0)}=0$. Thus, a total degree of squeezing $e^{2r}$ is achieved after a  time $\tau=\pi/2\omega_1+\pi/2\omega_2$.

\section{Connection between the matrix and the Gaussian formalism}\label{apd:sec:matrix_to_Gaussian_formalism}

The Gaussian process described by \cref{equ:full_master_equation} can be fully characterised by the first and second moments of the quadratures by taking $\df \braket{\hO} = \tr{\hO\,\df \hr}$ and using the invariance of trace under cyclic permutation of its arguments~\cite{Doherty_1999}. Given the definitions $\hat{\bm{r}} = \{\hx, \hp\}$ and $\bm{\sigma}_{\bm{r}_k, \bm{r}_j}=\frac{1}{2}\left(\braket{\hat{\bm{r}}_k\hat{\bm{r}}_j}+\braket{\hat{\bm{r}}_j\hat{\bm{r}}_k}\right)-\braket{\hat{\bm{r}}_k}\braket{\hat{\bm{r}}_j}$, we can write the equation for the system variances as
\begin{subequations}\label[equation]{apd:equ:variance_dynamics}
    \begin{align}
        \dot{\sigma}_{x,x} &= \frac{2}{m} \sigma_{x,p} - \lambda\sigma_{x,x} + \frac{\gamma\hbar^2}{8 k_B m T_{\text{cl}}} + \frac{\lambda\hbar}{2 m \omega}(2\overline{n}+1)  \notag\\
        & - 8\eta \Lambda \sigma^2_{x,x},\\
        \dot{\sigma}_{p,p} &= -2m\omega^2\sigma_{x,p} + (2\gamma-\lambda)\sigma_{p,p} + 2 \gamma k_B m T_{\text{cl}} \notag\\
        &+ \frac{1}{2}\lambda\hbar m\omega (2\overline{n}+1)+ 2\Lambda \hbar^2 - 8\eta\Lambda \sigma^2_{x,p},\\
        \dot{\sigma}_{x,p} &= \frac{1}{m}\sigma_{p,p} - m \omega^2\sigma_{x,x} -(\lambda-\gamma)\sigma_{x,p} 
         - \frac{i\hbar}{2}(\lambda+\gamma)\notag\\ 
         &- 8\eta \Lambda \sigma_{x,x}\sigma_{x,p}.
    \end{align}
\end{subequations}

The original  Riccati equation  \cref{equ:riccati_equation} reads \cite{Genoni_2016}
\begin{subequations}
    \begin{align}
        \dot{\bm{\sigma}} &= \Tilde{\bm{A}}\bm{\sigma} + \bm{\sigma} \Tilde{\bm{A}}^T + \Tilde{\bm{D}} - \bm{\sigma}\bm{B} \bm{B}^T\bm{\sigma}, \label{apd:equ:riccati_equation}\\
        \intertext{where}
        \Tilde{\bm{A}} &= \bm{B} - \bm{\Omega} \bm{C} \bm{\sigma}_B \frac{1}{\bm{\sigma}_B + \bm{\sigma}_M} \bm{\Omega} \bm{C}^T, \\
        \Tilde{\bm{D}} &= \bm{D} + \bm{\Omega} \bm{C} \bm{\sigma}_B \frac{1}{\bm{\sigma}_B + \bm{\sigma}_M} \bm{\sigma}_B \bm{C}^T \bm{\Omega}, \\
        \bm{B} &= \bm{C}\bm{\Omega} \sqrt{\frac{1}{\bm{\sigma}_B + \bm{\sigma}_M}},\\[2ex]
        \bm{\Omega}&=
        \begin{pmatrix}
            0 & 1\\
            -1 & 0
        \end{pmatrix}
        ,\quad
        \bm{C}=2\sqrt{\Lambda}~
        \begin{pmatrix}
            1 & 0\\
            0 & 0
        \end{pmatrix},\label{apd:equ:symplectic_matrix}
    \end{align}
\end{subequations}
with $\bm{A}$ being the drift matrix, $\bm{D}$ the diffusion matrix, $\bm{\Omega}$ the symplectic matrix, $\bm{C}$ the coupling matrix between the system and the light mode, $\bm{\sigma}_B$ and $\bm{\sigma}_M$ the CMs for the light mode and the measurement, respectively. When the continuous measurement is considered, and assuming no thermal photon, the CM for the light mode is
\begin{equation}\label{equ:bath_matrice}
    \bm{\sigma}_B =
    \left(\bar{n}+\frac{1}{2}\right)\iden_2,
\end{equation}
where $\iden_2$ is the identity matrix. 
For a perfectly efficient measurement ($\eta=1$), the CM for the Gaussian measurement reads
\begin{equation}\label{equ:pure_gaussian_measure}
    \bm{\sigma}_M =
    \begin{pmatrix}
        s & 0 \\
        0 & 1/s
    \end{pmatrix}\bm{\sigma}_B.
\end{equation}
Here, the factor $s\in (0,\infty)$ characterises the type of the general-dyne detection: the choice $s=1$ ($s\rightarrow \infty$) is for a heterodyne (detection and  gives the (homodyne) measurement.
For an inefficient Gaussian measurement ($\eta\in]0,1[$), the resulting CM is that of a mixture \cite{Genoni_2016,Shackerley-Bennett_2018} 
\begin{equation}\label{equ:mixed_gaussian_measure}
    \bm{\sigma}_M^\ast = \frac{1}{\eta}\bm{\sigma}_M  + \frac{1-\eta}{\eta}\bm{\sigma}_B.
\end{equation} 
Considering the homodyne detection, and taking $\bm{\sigma}_M^\ast$ instead of $\bm{\sigma}_M$ in \cref{equ:riccati_equation}, one has
\begin{equation}\label{apd:equ:riccati_equation_settings_2}
    \tilde{\bm{A}}=\bm{A},\quad \tilde{\bm{D}}=\bm{D},\quad \bm{B}=
    \begin{pmatrix}
        0&b\\
        0&0
    \end{pmatrix}
\end{equation}
with $b=\sqrt{8\eta \Lambda/(2\bar{n}+1)}$. By comparing \cref{equ:riccati_equation,apd:equ:variance_dynamics,apd:equ:riccati_equation_settings_2}, one can get the values for $a_1,a_2,d_1,d_2, b$ in \cref{equ:riccati_equation_settings} that link \cref{equ:riccati_equation} to the stochastic master equation \cref{equ:full_master_equation} in the main text. This 
reduces to the Lyapunov equation in \cref{equ:riccati_equation} for $\eta=0$.
 
\section{Investigation on the squeezing protocol in open dynamics}\label{apd:sec:diffusion_equation_transform}
Here we derive the solution of \cref{equ:riccati_equation} following the method in Ref.~\cite{Ntogramatzidis_2011,Behr_2019}.

\noindent
\emph{Quantum Lyapunov equation -- } 
For $\eta=0$, \cref{equ:riccati_equation} can be transformed to a first-order linear differential equation
\begin{equation}\label{apd:equ:linear_differentical_equation}
    \frac{\df }{\df t}
    \begin{pmatrix}
        \bm{N}_t\\
        \bm{\Psi}_t
    \end{pmatrix} 
    = \bm{\calH}
    \begin{pmatrix}
        \bm{N}_t\\
        \bm{\Psi}_t
    \end{pmatrix},
    \text{ with }
    \bm{\calH} = 
    \begin{pmatrix}
        -\bm{A}^T & \bm{0} \\
        \bm{D}    & \bm{A}
    \end{pmatrix},
\end{equation}
where we introduce
\begin{equation}\label{apd:equ:new_variables}
    \bm{\Psi}_t =\bm{\sigma}_t \bm{N}_t, \quad\text{and}\quad \dot{\bm{N}}_t=-\bm{A}^T \bm{N}_t,
\end{equation}
and let $\bm{N}_0$ to be the identity matrix $\bm{\iden}$. The linear different equation given by \cref{apd:equ:linear_differentical_equation} can be solved by making the matrix $\bm{\calH}$ diagonal. Consider the similarity transformation 
\begin{equation}
    \bm{T}=
    \begin{pmatrix}
        \iden & \bm{0} \\
        \bm{X} & \iden
    \end{pmatrix},
\end{equation}
such that $\bm{T} \bm{T}^{-1}=\bm{T}^{-1} \bm{T} =\iden$, and define $\bm{X}$ as the solution to the equation 
\begin{equation}\label{apd:equ:characteristic_function}
    \bm{A} \bm{X} + \bm{X} \bm{A}^T + \bm{D} = \bm{0}.
\end{equation}
Apply the transformation on $\bm{\calH}$, such that
\begin{equation}
    \Tilde{\bm{\calH}}\equiv \bm{T}^{-1}{\bm{\calH}}\bm{T} = 
    \begin{pmatrix}
        -\bm{A}^T & \bm{0} \\
        \bm{0} & \bm{A}
    \end{pmatrix}.
\end{equation}
where the transformed matrix $\bm{\calH}$ becomes diagonal. Correspondingly, \cref{apd:equ:linear_differentical_equation} is transformed to
\begin{equation}\label{apd:equ:transformed_linear_differentical_equation}
    \frac{\df }{\df t}
    \begin{pmatrix}
        \Tilde{\bm{N}}_t \\
        \Tilde{\bm{\Psi}}_t
    \end{pmatrix}
    =\tilde{\bm{\calH}}
    \begin{pmatrix}
        \tilde{\bm{N}}_t \\
        \tilde{\bm{\Psi}}_t
    \end{pmatrix},
\end{equation}
with the transformed vector defined as
\begin{equation}
    \begin{pmatrix}
        \Tilde{\bm{N}}_t \\
        \Tilde{\bm{\Psi}}_t
    \end{pmatrix}
    \equiv \bm{T}^{-1}
    \begin{pmatrix}
        \bm{N}_t \\
        \bm{\Psi}_t
    \end{pmatrix}
    =
    \begin{pmatrix}
        \bm{N}_t \\
        \bm{\Psi}_t - \bm{X} \bm{N}_t
    \end{pmatrix}.
\end{equation}
The transformed differential equation has the solution
\begin{equation}
    \begin{pmatrix}
        \Tilde{\bm{N}}_t \\
        \Tilde{\bm{\Psi}}_t
    \end{pmatrix}
    = e^{t\Tilde{\bm{\calH}}}
    \begin{pmatrix}
        \Tilde{\bm{N}}_0 \\
        \Tilde{\bm{\Psi}}_0
    \end{pmatrix},
\end{equation}
which in details reads
\begin{equation}\label{apd:equ:new_variables_solutions}
    \begin{aligned}
        \bm{N}_t &= \Tilde{\bm{N}}_t = e^{-t\bm{A}^T}\Tilde{\bm{N}}_0= e^{-t\bm{A}^T}\bm{X}_0= e^{-t\bm{A}^T}, \\
        \bm{\Psi}_t &-\bm{X}\bm{N}_t= \Tilde{\bm{\Psi}}_t = e^{t\bm{A}}\Tilde{\bm{\Psi}}_0= e^{t\bm{A}}(\bm{\sigma}_0-\bm{X}).
    \end{aligned}
\end{equation}
By applying \cref{apd:equ:new_variables} to \cref{apd:equ:new_variables_solutions}, one can get the solution to the CM for the system $\bm{\sigma}_t$ as in \cref{equ:langevin_equation_solution} for the quantum Lyapunov equation.
If the squeezing protocol is not performed, \ie a time-independent QHO with constant drifting and diffusion matrices $\bm{A}$ and $\bm{D}$, the asymptotic state for \cref{equ:langevin_equation_solution} can be easily shown to be the characteristic matrix $\bm{X}$ given by \cref{apd:equ:characteristic_function}, such that $\bm{\sigma}_{t\to\infty} = \bm{X}$. 

One can attempt to find the asymptotic state of the squeezing dynamics described by \cref{equ:langevin_squeezing}. By introducing $\Delta \bm{X} = \bm{X}_1 - \bm{X}_2$ and $\bm{\alpha}$ satisfying 
\begin{equation}
    \bm{\alpha} - e^{t_2 \bm{A}_2}e^{t_1 \bm{A}_1}\bm{\alpha} e^{t_1 \bm{A}_1^T} e^{t_2 \bm{A}_2^T} = \Delta\bm{X} - e^{t_2 \bm{A}_2}\Delta \bm{X} e^{t_2 \bm{A}_2^T}, \label{equ:langevin_alpha}
\end{equation}
the CM for one round of squeezing, i.e. at time $\tau = t_1+t_2$, results from 
\begin{equation}
    \bm{\sigma}_{\tau} - \bm{X}_1 + \bm{\alpha} = e^{t_2\bm{A}_2}e^{t_1\bm{A}_1}(\bm{\sigma}_0-\bm{X}_1 + \bm{\alpha})e^{t_1\bm{A}_1^T}e^{t_2\bm{A}_2^T},
\end{equation}
and the asymptotic state is given by $\bm{\sigma}^\text{sq}_{t\to\infty} = \bm{X}_1 - \bm{\alpha}$.

\noindent
\emph{Quantum Riccati equation -- } 
A similar transformation can be done introducing
\begin{equation}
    \bm{\Psi}_t =\bm{\sigma}_t \bm{N}_t, \quad\text{and}\quad \dot{\bm{N}}_t=-\bm{A}^T \bm{N}_t + \bm{B} \bm{B}^T \bm{\sigma}_t \bm{N}_t,
\end{equation}
such that one has
\begin{equation}\label{apd:equ:linear_differential_equation_riccati}
    \frac{\df }{\df t}
    \begin{pmatrix}
        N_t\\
        \Psi_t
    \end{pmatrix} 
    = \calH
    \begin{pmatrix}
        N_t\\
        \Psi_t
    \end{pmatrix},
    \text{ with }
    \calH = 
    \begin{pmatrix}
        -A^T & BB^T \\
        D    & A
    \end{pmatrix},
\end{equation}
and $\bm{N}_0$ is given by the identity matrix. Introduce two similarity transformations,
\begin{equation}
    \bm{T}_1=
    \begin{pmatrix}
        \iden & \bm{0} \\
        \bm{{\cal X}}_1 & \iden
    \end{pmatrix},
    \quad
    \bm{T}_2=
    \begin{pmatrix}
        \iden & \bm{{\cal X}}_2 \\
        \bm{0} & \iden
    \end{pmatrix},
\end{equation}
and letting $\bm{{\cal X}}_1$ and $\bm{{\cal X}}_2$ be the solutions to the equations 
\begin{subequations}
    \begin{align}
        &\bm{A} \bm{{\cal X}}_1 + \bm{{\cal X}}_1 \bm{A}^T + \bm{D} - \bm{{\cal X}}_1\bm{B}\bm{B}^T\bm{{\cal X}}_1 = \bm{0},\\
        &\bm{{\cal A}}^T\bm{{\cal X}}_2 + \bm{{\cal X}}_2\bm{{\cal A}} - \bm{B}\bm{B}^T = \bm{0},
    \end{align}
\end{subequations}
the characterisation matrix is diagonalised such that
\begin{equation} 
    \bm{\calH}'' = \bm{T}_2^{-1}\bm{T}_1^{-1}\bm{{\cal H}}\bm{T}_1\bm{T}_2=
    \begin{pmatrix}
        -\bm{{\cal A}}^T & \bm{0} \\
        \bm{0} & \bm{{\cal A}}
    \end{pmatrix},
\end{equation}
where we define $\bm{{\cal A}}=\bm{A}-\bm{{\cal X}}_1\bm{B}\bm{B}^T$. Define the transformed vector 
\begin{align}
    \begin{pmatrix}
        \bm{N}''_t \\
        \bm{\Psi}''_t
    \end{pmatrix}
    & \equiv \bm{T}_2^{-1}\bm{T}_1^{-1}
    \begin{pmatrix}
        \bm{N}_t \\
        \bm{\Psi}_t
    \end{pmatrix}\\
    & =
    \begin{pmatrix}
        \bm{N}_t + \bm{{\cal X}}_2\bm{{\cal X}}_1\bm{N}_t-\bm{{\cal X}}_2\bm{\Psi}_t \\
        \bm{\Psi}_t - \bm{{\cal X}}_1 \bm{N}_t
    \end{pmatrix},
\end{align}
The differential equation in \cref{apd:equ:linear_differential_equation_riccati} can be solved as
\begin{equation}
    \begin{pmatrix}
        \bm{N}''_t \\
        \bm{\Psi}''_t
    \end{pmatrix}
    = e^{t\bm{\calH}''}
    \begin{pmatrix}
        \bm{N}''_0 \\
        \bm{\Psi}''_0
    \end{pmatrix},
\end{equation}
and the solution on the CM for the system is then given by
\begin{equation}\label{apd:equ:riccati_solution}
    (\bm{\sigma}_t-\bm{{\cal X}}_1)\bm{\xi}_t^{-1}=e^{t\bm{{\cal A}}}(\bm{\sigma}_0-\bm{{\cal X}}_1)\bm{\xi}_0^{-1}e^{t\bm{{\cal A}}^T},
\end{equation}
where we take $\bm{\xi}_t = \iden -\bm{{\cal X}}_2(\bm{\sigma}_t-\bm{{\cal X}}_1)$. Take the inverse of \cref{apd:equ:riccati_solution} and define $\bm{\delta}_t = (\bm{\sigma}_t-\bm{{\cal X}}_1)^{-1}$, the solution gives \cref{equ:riccati_equation_solution} in the main text.

The asymptotic state of \cref{equ:riccati_equation_solution} without the squeezing protocol reads  $\bm{\sigma}_{t\to\infty} = \bm{X}_1 + \bm{X}_2^{-1}$. The asymptotic state with the squeezing protocol is however difficult to attain, therefore we take the numerical solution instead.

\section{The squeezing protocol}\label{apd:sec:squeezing_paramters}

Suppose the quadrature variance follows the relation $\sigma_{n+1} = e^{sr}\sigma_n + \chi$ after one squeezing process, which can be recast into the form 
\begin{equation}
 \sigma_{n+1} - \frac{\chi}{1-e^{sr}} = e^{sr}\left(\sigma_n- \frac{\chi}{1-e^{sr}}\right).   
\end{equation}
One can see that, for a finite $\chi>0$, this dynamics leads the variance to a steady positive value $\sigma_\infty \to \frac{\chi}{1-e^{sr}}$ when ${sr}<0$, or to the infinity $\sigma_\infty \to \infty$ when ${sr}>0$.
The squeezing parameters in our protocol can be acquired by setting the diffusion to be zero in \cref{equ:langevin_squeezing} ($d_{1,2}\to0$ leads to $\chi\to0$ while leaving $e^{sr}$ unchanged), which read
\begin{align}
{sr}_{xx} &= f + \ln\left({\Omega_1^2}/{\Omega_2^2}\right), \\
{sr}_{pp} &= f + \ln\left(\frac{\Omega_2^2}{\Omega_1^2} + \Delta a \frac{m (\Omega_2^2-\Omega_1^2)}{\Omega_1^2}\frac{\sigma_{xp,n}}{\sigma_{pp,n}}\right. \notag\\
& ~~~~~~~~+\left. \Delta a^2 \frac{m^2 (\Omega_2^2-\Omega_1^2)^2}{4\Omega_1^2\Omega_2^2}\frac{\sigma_{xx,n}}{\sigma_{pp,n}} \right),\\
{sr}_{xp} &= f + \ln\left(1  + \Delta a \frac{m (\Omega_2^2-\Omega_1^2)}{2\Omega_2^2}\frac{\sigma_{xx,n}}{\sigma_{xp,n}} \right),
\end{align}
where $f={\pi(a_1+a_2)(\Omega_1+\Omega_2)}/{(2\Omega_1\Omega_2)}$ and $\Delta a = a_2-a_1$. In the limit of $\Delta a\to 0$, this gives \cref{equ:momentum_squeezing_ratio}.

\section{Range of coefficients}\label{apd:sec:range_of_coefficients}

\begin{table}[tb]
    \centering
    \begin{tabular}{|c|c|c|c|c|c|c|c|c|c|}
        \hline
        Pars (unit)  & Values & Pars (unit)  & Values \\
        \hline
        $R$ (\si{\nm}) & 50 & $\lambda$ (\si{\nm}) & $1550$ \\
        \hline
        $P_t$ (\si{\W}) & 0.50 & $W_t$ (\si{\nm}) & $1000$ \\
        \hline
        $\varepsilon_0$ (\si{\F/\nm}) & $8.9\times 10^{-21}$ & $\epsilon$ & $2$ \\
        \hline
        $A_x$ & $1$ & $A_y$ & $0.9$\\
        \hline
    \end{tabular}
    \caption{Collection of the experimental parameters and their values for the computation of the photon-recoil rate.}
    \label{apd:tab:parameter_values}
\end{table}

The dissipation due to gas is described by \cref{equ:caldeira_leggett_dissipator} where $T$ is the temperature of the chamber and $\gamma$ is the viscous friction that can be calculated from kinetic gas theory \cite{Gieseler_2013,Chang1005}:
\begin{equation}
\gamma=\frac{64}{3} \frac{R^2P}{m v_\mathrm{gas}} \sim 10^3 (P/\si{\mbar})~\si{\Hz}.
\end{equation}
with the mean gas velocity $v_\mathrm{gas}=\sqrt{8k_bT_\text{cl}/(\pi m_\mathrm{gas})}$, and $m_\mathrm{gas}\approx 10^{-24}\,\si{\kg}$ is the average of the gas molecules. $R$ is the radius of the nanoparticle and $P$ is the pressure in the vacuum chamber. Our estimation has been achieved in the latest experiment \cite{Dania_2023}.
The thermal dissipation due to energy exchange with the laser is described by \cref{equ:thermal_dissipator}, where the coupling rate is estimated by $\lambda\sim\gamma$ as they contribute equally to the damping rate in momentum, \ie $a_2\approx \lambda + \gamma$. We set $\overline{n} = (\exp(\hbar\omega/k_bT) -1))^{-1}\sim 10^7$ as the mean occupation of the nanoparticle at $\omega/2\pi = 100$\,\si{\kHz} and $T = 50$\,\si{\K}. This gives the estimation of the thermalisation rate such that
\begin{equation}
    \lambda \overline{n} \sim 10^{8} (P/\si{\mbar})~\si{\Hz}.
\end{equation}
The position detection due to the measurements of the photons scattered back from the nanoparticle is described by \cref{equ:local_dissipator}, where the coupling strengths reads \cite{Gardiner_2004,Gonzalez-Ballestero_2019,Seberson_2020}
\begin{equation}\label{apd:equ:photon_recoil_rate}
    \Lambda=\frac{7 \pi \varepsilon_0}{30\hbar} \left(\frac{\epsilon_c V E_t}{2\pi}\right)^2 k_0^5 \sim 7 \times 10^{25}~(\si{\m^{-2}\Hz}),
\end{equation}
where $\varepsilon_0$ is the vacuum permittivity, $\epsilon_c=3{(\varepsilon-1)}/{(\varepsilon+2)}$, and $\epsilon$ is the relative dielectric constant of the nanoparticle. $V$ is the volume of the nanoparticle, $k_0=\omega_0/c$ with $c$ being the speed of light and $\omega_0 = {2\pi c}/{\lambda}$ the laser beam frequency. We take $E_t=\sqrt{{4P_t}/{\pi\varepsilon_0 c W_t^2A_xA_y}}$ where $P_t$ is the tweezer power, $W_t$ is the tweezer waist, $A_x$ and $A_y$ are the asymmetry factor \cite{Gonzalez-Ballestero_2019}. Taking the values in \cref{apd:tab:parameter_values}, we give the estimation of the rate for the photon recoils to be $\Lambda \sim 10^{26}$\,\si{\m^{-2}\Hz}. Reducing the photon-recoil rate by $3$ orders seems to be possible in a future experiment, by reducing the tweezer power to $30\,\si{\mW}$ \cite{Kustura_2022}, the particle radius to $40\,\si{\nm}$ and the tweezer waist to $2\,\si{\um}$ (or just the particle radius to $35$\,\si{\nm}). Based on these estimation we provide our discussion below \cref{equ:photon_recoil_rate}.

\end{document}